\documentclass[]{article}

\usepackage[margin=1in]{geometry}
\usepackage[utf8]{inputenc}
\usepackage{mathtools}

\title{Using fuel cells to power electric propulsion systems}

\author{
Lui Habl\thanks{MSc student, Physics institute, lui.habl@aerospace.unb.br.}%
}

\date{}

\begin{document}

\maketitle

\begin{abstract}
The origin of fuel cell technology has a notable connection to the history of spaceflight, having been used in remarkable programs such as Gemini, Apollo and the Space Shuttle. With the constant growth of the electric propulsion technology in the last years, one natural application for fuel cells to be considered would be the electrical feeding of those thrusters for different mission profiles. In this article we explore in details this possibility, showing what would the necessary characteristics of such a device be in order to improve mission parameters, as payload ratio and thrust, among others. In the first section, a brief review of the applications of fuel cells in the space industry is shown, and the classical analytical modeling of these devices is presented. In the second part, two case studies illustrating most of the possible ways to use fuel cells in conjunction with electric propulsion systems are shown , and the possible advantages and limitations of the applications are demonstrated analytically. The result of the analysis shows that, in the case where the fuel cell reaction products are disposed of and the propulsion has its own feed system, the application of fuel cell technology would bring no advantages for most kinds of missions. On the other hand, when the case where the fuel cell exhaust is used as a propellant is considered, it is shown that it is possible to improve mission parameters, such as thrust and specific impulse, if certain conditions are met.  
\end{abstract}

\section{Introduction}

The birth of the modern fuel cell concept dates back to 1838, when the Welsh scientist Sir William Grove first proposed and demonstrated what is known today as the Grove cell \cite{app90}. Despite the fact that research in this field continued to be developed through the years, just in the 1960s, with the work of the chemists Thomas Grubb and Leonard Niedrach of General Electric (GE) inventing the polymer electrolyte membrane (PEM) fuel cells, the first commercial systems were developed, finding their application on the Gemini missions. The space and fuel cell technologies remained connected until nowadays, and, for example, missions as the Apollo and the Space Shuttle employed hydrogen-oxygen alkali fuel cells (AFC), which were the only power system type available at that time that was able to meet the necessary energy requirements \cite{men08}. With the involvement of the aerospace, energy and automotive industries, fuel cell systems achieved a high development level,  being used today in a large range of applications, in situations where electrochemical conversion is deemed convenient. 

In general, the operation of a fuel cell follows a scheme where the reactants are injected on the outer surface of the anode and cathode, which are composed by a catalyst layer. At the anode, the fuel is oxidized producing electrons, while ions, through diffusion, migrate in the electrolyte to the cathode. The electrons are provided to the load, creating the useful current, and come back to the cathode catalytic layer to recombine with the oxidizer, completing the so-called cathodic oxidation-reduction reaction. As the voltage of a single fuel cell is always limited by the electrochemical potential of the reaction, usually smaller than 1V, it is necessary to make different arrangements to guarantee the necessary output voltage level. The series, or series-parallel, combination is called a fuel cell stack. On the other hand, the current, and therefore the power, level is solely determined by the area of the exposed electrodes and the reactants mass flow rate provided to the cells. 

Through the years, many different classes of fuel cell systems were developed, with diverse unique characteristics, and most of them being categorized by the electrolyte material employed \cite{men08}. The most popular types are described below \cite{spi07}:

\begin{itemize}
\item Alkaline fuel cell (AFC): is a low temperature fuel cell, usually operating between 60-250 ºC. Has its electrolyte composed usually of a solution of potassium hydroxide and water, which is highly sensible to contamination by $CO_2$. To avoid poisoning, the AFC must run on highly pure oxygen, thus being adequate mostly to space applications. On the other hand, this type has a high efficiency when compared to others with minimum reaction losses. 
\item Solid oxide fuel cell (SOFC): is a high temperature fuel cell, operating between 600-1000 ºC. Its electrolyte is composed of Yttria and stabilized Zirconia, with its major poison being sulfur. This material is very resistant to $CO$, thus this type of fuel cell is very tolerant to a variety of different fuels. The main disadvantage, besides the high operational temperature, is the long startup time, which limits the application of this fuel cell type on systems with intermittent operation. 
\item Phosphoric acid fuel cell (PAFC): is a medium temperature fuel cell, with its operational temperature ranging between 150-230 ºC. The electrolyte is composed of a solution of phosphoric acid in a silicon carbide matrix. Sulfur and $CO$ are the greatest sources of poisoning in this fuel cell, still, in normal operational mode the electrolyte can withstand up to 2.5\% of $CO$ contamination, allowing more flexible applications. These cells shown a high capability of heat wasting and a considerable durability, thus being mostly used in stationary power sources.  
\item Molten carbonate fuel cell (MCFC): is also a high temperature system, operating in the 600-800 ºC range. Its electrolyte is usually molten alkali metal carbonates in a highly porous matrix, with its major poison being sulfur. This material is very tolerant to $CO$ poisoning and also fuel flexible, thus allowing a wide range of applications. The major disadvantage is that these systems have usually very high startup time, permitting only continuous power applications.
\item Polymer electrolyte membrane (PEM) fuel cell: is a very low temperature cell, having its operational temperature in the 30-100 ºC range. Its electrolyte is composed of a polymeric membrane of solid perfluorosulfonic acid, with its major poison being $CO$, sulfur, metal ions and peroxide. Despite being very sensible to several substances, this cell have very high performance levels, with an elevated power density and short startup time. Because of these characteristics, and also its portability, this system is one of the most used nowadays, especially in the automotive industry.
\end{itemize}

In parallel, during the last 50 years electric propulsion systems has been getting a considerable role in the space technology community with applications ranging from interplanetary probes, as the SMART-1 \cite{kop05} and Deep Space 1 \cite{bro02}, to orbit raising and station keeping of telecommunications satellites \cite{maz16}\cite{cas15}. Several different types of thrusters were proposed, including \cite{maz16}: Pulsed Plasma Thrusters (PPTs), Magnetoplasmadynamic Thrusters (MPD-T), Field Effect Emission Propulsion (FEEP), and the two most employed nowadays, Hall Effect Thrusters (HET), and Gridded Ion Engines (GIE). Each of these systems operate in a different range of thrust-to-power ratio, thus being ideal to different mission profiles \cite{tur06}.

After citing both systems, one natural proposition would be their combination in order to improve the power output temporarily to the thruster, thus creating a chemical or solar-chemical electric propulsion system, allowing different mission profiles that need a short high power impulse using an electric thruster. Some examples would be fast orbit raising for GEO satellites, attitude control and rendezvous maneuvers. 

There are several possible ways to arrange both systems and it is possible classify these different cases depending on two basic characteristics: (1) whether the fuel cell output is re-used as a propellant; and (2) whether the power used by the thruster comes just from the fuel cell or from both the fuel cell and the solar arrays. With this, four different architectures must be studied in order to completely assess the feasibility of combination of the systems.

In this article we explore in details this possibility, showing what would be the necessary characteristics of the device the make mission parameters (payload ratio, thrust, among others) improve. In the first section, it is shown a brief review of the applications of fuel cells in the space industry, and it is presented the classical analytical modeling of these devices. In the second part, it is shown two case studies showing most of the possible ways to use fuel cells in conjunction with electric propulsion systems, and it is demonstrated analytically the possible advantages and limitations of the applications. 

\section{Fuel cell modeling}

The modeling of fuel cells is fundamentally a complex electrochemical question and a precise computation of their performance is not necessary for the development of this study. Mench \cite{men08} and Spiegel \cite{spi07} present a broad review about the modeling and construction of these systems. The simplified model presented below is based on these two works.

In order to compute the approximated performance of FCs it is necessary to take into consideration the Faraday conservation laws, and it is possible to show that mass output of a fuel cell is given by,

\begin{equation}
\dot{m}_{FC} = \frac{I}{n F} M_{FC}.
\end{equation}

Where $I$ is the produced current, $n$ is the number of equivalent electrons produced per mole of reaction, $M_{FC}$ is the molar mass of the product and $F$ is the Faraday constant. Considering that the produced power is simply given by $P_{FC}=E_c I$, where $E_c$  is the voltage generated in a single cell, it is possible to determine a direct relation between power and mass flow rate as,

\begin{equation}
\frac{P_{FC}}{\dot{m}_{FC}} = \frac{E_c n F}{M_{FC}}.
\end{equation}

Using the definition of the Gibbs energy of formation as being $\Delta G^0=-E^0 nF$, where $E^0$ is the maximum achievable voltage, it is possible to define the specific energy, in J/kg, generated by the used substances,

\begin{equation}
k_c = - \eta_v \frac{\Delta G^0}{M_{FC}},
\end{equation}

where $\eta_v=E_c/E^0$ is the voltage efficiency of the fuel cell.

\begin{table}
\centering
\caption{Specific energy values of possible reactions considering no losses}
\label{t1}
\begin{tabular}{@{}lll@{}}
\hline
\textbf{Reaction}                    										 & \textbf{Specific Energy ($\eta_v = 1$)} & \textbf{Technology readiness}      \\ \hline
Hydrogen ($H_2$) – Oxygen ($O_2$)             			& 13.16 MJ/kg                       & Commercial\cite{men08}                         \\
Methanol ($CH_3OH$) – Oxygen ($O_2$)             & 8.77 MJ/kg                       & Commercial\cite{liu2006} \\
Hydrazine ($N_2H_4$) – Hydrogen Peroxide ($H_2O_2$) & 8.86 MJ/kg                       & Tested \cite{ser2010}                            \\
MMH ($CH_3(NH)NH_2$) – NTO  ($N_2O_4$)                            & 8.74 MJ/kg                       & Not tested                         \\
Ammonia ($NH_3$) – Oxygen ($O_2$)              & 8.27 MJ/kg                       & Tested \cite{maf2005}                         \\ \hline

\end{tabular}
\end{table}

\section{Study of cases}
In order to assess the possible application of fuel cells to power up electric thrusters, two basic architectures will be studied next. The first case consists on using of a fuel cell as a power source alongside with the solar cells, with all its reaction products being rejected in a zero-velocity exhaust producing no thrust. In this architecture it would be also possible to consider a finite specific impulse for the fuel cell output if, for example, it was used as a working fluid for a cold gas thruster in order to improve performance. However, for the sake of simplicity and since the impact of a low specific impulse device would have on the general performance of an electric propulsion system is almost negligible, we only consider here the limit case where the exhaust velocity tends toward zero.

In the second case, the fuel cell is again used as a source in conjunction with the spacecraft solar cells, however, in this scenario the reaction products are used as the propellant for the thruster

\subsection{Case 1: separate propellant feeding system}
In order to compute the overall performance of the spacecraft it is necessary to define first the basic energetic relations between the propulsion and the FC systems. The power output of the FC, $P_{FC}$, is considered to be proportional to a generic mass flow rate input, $\dot{m}_{FC}$, in the form,

\begin{equation}
P_{FC} = k_c \dot{m}_{FC},
\end{equation}

where $k_c$  is a conversion factor in J/kg and also represents the specific energy of the FC as shown in the first section. 

If we assume then that a thruster of constant efficiency $\eta_T$ is used, and that it receives a total power $P_T = P_{SC} + P_{FC}$, where Psc is the power generated by the solar cells, the resulting thrust, $T$, can be computed as,

\begin{equation} 
\label{e5}
T =\frac{2\eta_T}{u_e} (P_{SC} + k_c \dot{m}_{FC}),
\end{equation}

where $u_e$ is the exhaust velocity of the thruster in m/s. Furthermore, for the sake of simplicity it is then defined that the total power provided to the thruster $P_T$ is a multiple of the spacecraft solar cell power in the form,

\begin{equation}
\label{e6}
P_T = \varepsilon P_{SC} = P_{SC} + k_c \dot{m}_{FC}.
\end{equation}

Of course, simultaneously, equation \ref{e5} and \ref{e6} also defines that the resultant thrust will also be a multiple of the thrust given by the solar power in the form $T=\varepsilon T_{SC}$.

Using the relation above, the FC mass flow rate is then simply,

\begin{equation}
\label{e7}
\dot{m}_{FC} = \frac{\varepsilon - 1}{k_c} P_{SC}.
\end{equation}

If it is then defined that the thrust is also given by $T=\dot{m}_{p} u_e$, where $\dot{m}_{p}$ is the propellant mass flow rate in the thruster, and the equations \ref{e5} and \ref{e7} are taken into consideration, it is possible to relate all major quantities as,

\begin{equation}
\frac{\dot{m}_{FC}}{u_e^2} = \left( \frac{\varepsilon - 1}{\varepsilon}\right)\frac{\dot{m}_{p}}{2 \eta_T k_c}.
\end{equation}

Noting that all the parameters in the right hand side, except $\varepsilon$, can be taken as constant, it is possible to see that this equation shows a direct relation between the three major quantities in the model: power, fuel cell mass flow rate and the generated specific impulse. Another consequence of this equation is that when $\varepsilon \gg 1$ the left hand side ratio tends towards a fixed value of $\dot{m}_{p}/2\eta_T k_c$ (independent of $\varepsilon$), as the solar power starts to be insignificant compared to the power generated by the FC.

To account for the influence of the power system input on the mission effectiveness itself, it is possible to define a modified thrust equation that takes into consideration the additional propellant consumption. If it is considered that all the mass consumed by the fuel cell is disposed as a thruster exhaust with negligible specific impulse, the equivalent thrust is,

\begin{equation}
T_q = u_q (\dot{m}_p + \dot{m}_{FC}) = u_e \dot{m}_p,
\end{equation}

where $u_q$ is defined as an equivalent exhaust velocity, that represents the combined system as a single thruster. As shown in the relation, this equivalent exhaust velocity is defined as a ratio of propellant mas flow rates times the real exhaust velocity resultant from the power increment, $u_e$. In order improve the comprehensibility of the model, it is interesting to introduce a set of normalized variables, $\chi=u_q/u_{SEP}$, $\lambda=k_c \dot{m}_p/P_{SC}$ and $\nu=\Delta V/u_{SEP}$, where $u_{SEP}=\sqrt[]{2\eta_T P_{SC}/\dot{m}_p }$. Using equations 8 and 9, it is possible to show explicitly that,

\begin{equation}
\label{e10}
\chi = \frac{\lambda  \sqrt{\varepsilon}}{\lambda + \varepsilon - 1}.
\end{equation}

Using the obtained relation for exhaust velocity it is possible to redefine then the Tsiolkovsky equation for payload ratio,

\begin{equation}
\label{e11}
\frac{m_f}{m_0} = \exp{\left(-\frac{\Delta V}{u_q(\varepsilon)}\right)} = \exp {\left( - \nu \frac{\lambda + \varepsilon - 1}{\lambda  \sqrt{\varepsilon}} \right)},
\end{equation}

where $m_f$ and $m_0$ are respectively the dry and initial mass of the spacecraft.

Once that the $\Delta V$ is considered constant for the mission, the exponential function present in the last equation varies monotonically with $u_q (\varepsilon)$. With this assumption it is then possible to compute an extremum of the payload ratio by finding the extremum of $u_q$. This can be done simply finding an $\varepsilon$ such that,

\begin{equation}
\label{e12}
\frac{d\chi}{d\varepsilon} = \frac{\lambda (\lambda - \varepsilon - 1)}{2 \sqrt{\varepsilon} (\lambda + \varepsilon - 1)^2}.
\end{equation}

Since only the numerator can make the function go to zero, the extremum can be found at,

\begin{equation}
\label{e13}
\varepsilon_m = \lambda - 1.
\end{equation}

As it can be seen from equation \ref{e12}, when $\varepsilon<\varepsilon_m$, $u_q$ and the payload ratio will be in a region of constant increment and the increasing of power supplied by the FC will be advantageous in a efficiency point of view. Of course, it is also straightforward to imply that if $\varepsilon_m>1$ the application of the proposed system will always increase the payload fraction delivered to the final orbit. Hence, this allows to be define a basic requirement for a minimum performance of the FC that guarantees an increasing mass efficiency,

\begin{equation}
\label{e14}
k_c > 2\frac{P_{SC}}{\dot{m}_p} = \frac{u^2_{SEP}}{\eta_T}.
\end{equation}

\begin{figure}

  \centering
    \includegraphics[width=0.6\textwidth]{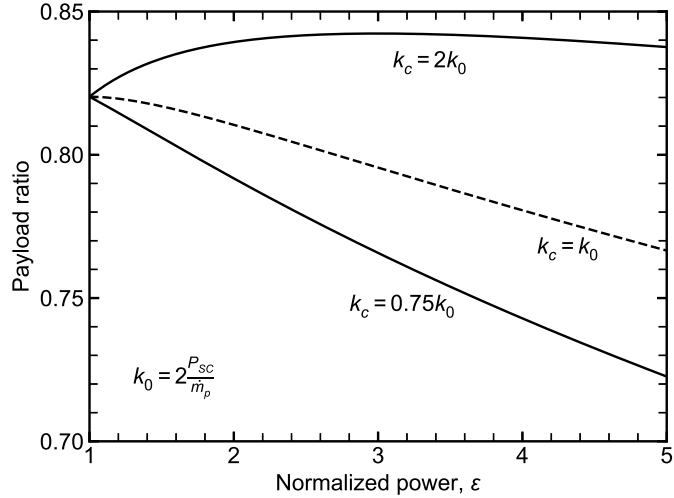}
  
  \caption{Variation of payload mass ratio with the normalized power. Three cases are shown, with a crescent profile $k_c>k_0$, the limit case with zero derivative at $\varepsilon=1$ $k_c=k_0$ and the pure decreasing case $k_c<k_0$. The profile assumes the performance of XIPS 25-cm average $I_{sp}$ of 3550 seconds, thrust of 166 mN, efficiency of 68,8\%, at an input power of 4300 W.}
\label{fig1}
\end{figure}

Figure \ref{fig1} shows the behavior of the payload mass ratio,$m_f/m_0$, as function of $\varepsilon$ for a generic mission ($\nu = 0.2$) employing an ion thruster with similar characteristics of the 25-cm XIPS thruster, developed by L3 Communications, that has an average $I_{sp}$ of 3550 seconds, thrust of 166 mN, efficiency of 68,8\%, at an input power of 4300 W \cite{goe09}. The figure demonstrates the three main behaviors of the equation 11 depending of the value of $k_c$ compared to $k_0=2P_{SC}/\dot{m}_p$, as predicted by the requirement in the expression \ref{e14}.

If the model is expressed in terms of absolute power, using again $\varepsilon = P_T/P_{SC}$, it is possible to consider the case where the FC is the only power source of the propulsion system, without any input from the spacecraft’s solar cells ($P_{SC}=0$). With this equation \ref{e6} becomes $P_T= k_c \dot{m}_{FC}$. Since equation \ref{e13} maximizes $\chi$ it is possible to say that, when $P_{SC}=0$, $u_q$ maximum will happen with $P_T=k_c \dot{m}_p$, which is the same of $\dot{m}_{FC}=\dot{m}_p$. Using the equation \ref{e10} one can find that the theoretical maximum for the equivalent exhaust velocity in the system without solar cell power input is,

\begin{equation}
u_{q,m} = \sqrt[]{\frac{\eta_T k_c}{ 2}}.
\label{eq:uqm}
\end{equation}

\begin{figure}

  \centering
    \includegraphics[width=0.6\textwidth]{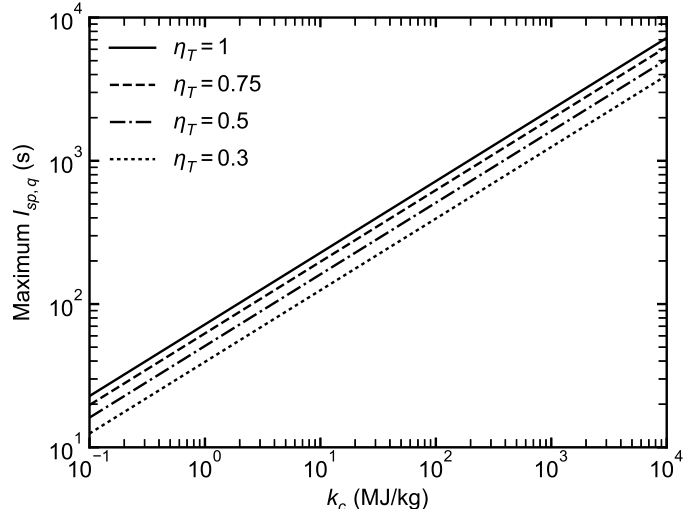}
  
  \caption{Maximum specific impulse for the case where the FC is the only power supply.}
\label{fig2}
\end{figure}

Figure \ref{fig2} plots the equivalent specific impulse, $I_{sp,q}=u_q/g_0$, for a range of specific energy values. It is possible to see that for all the real values of specific energy shown in table \ref{t1}, the attained equivalent specific impulse is considerably small when considering that this is an electric propulsion mission.

In fact, if it is taken into consideration that the theoretical maximum exhaust velocity obtained in a chemical thruster can be expressed as $u_{ch,m}=\sqrt{(-2\Delta H^0/M)}$ \cite{tur06}, where $\Delta H^0$ is standard enthalpy of formation and M is the exhaust molar mass of the propellant, it is possible to see that,

\begin{equation}
\frac{u_{ch,m}}{u_{q,m}} = 2  \sqrt{\frac{\Delta H^0 }{\Delta G^0 }}.
\end{equation}

Showing that, if $P_{SC}=0$, the maximum specific impulse obtained in the chemical thruster using the same fuel and oxidizer will always be higher, since $|\Delta G^0| \leq |\Delta H^0|$ for spontaneous reactions\cite{men08}. This is simply justified by the fact that in the chemical thruster the energy is converted directly and its exhaust is not disposed as in this study of case. 

With the general definitions, where the FC is the secondary supply, it is possible to compute the approximate time needed for the execution of a given maneuver. Considering that both the $\dot{m}_p$ and $\dot{m}_{FC}$ are constant throughout all the maneuver, the total mass of equivalent propellant used in the maneuver can be defined as $m_q=(\dot{m}_p+\dot{m}_{FC} )\Delta t$, where $\Delta t$ is the time in seconds, and considering $\tau=\Delta t \dot{m}_p/m_f$. Using again the simple Tsiolkovsky equation for propellant mass the approximate time for the maneuver is then,

\begin{equation}
\label{e18}
\tau = \frac{\lambda}{\lambda + \varepsilon - 1}\left( \exp\left( \nu \frac{\lambda + \varepsilon - 1}{\lambda \sqrt{\varepsilon}} \right)  - 1 \right).
\end{equation}

The analysis of the equation obtained for time is slightly more complex than the one for exhaust velocity and an analytical expression for the exact location of its extrema might be overcomplicated for the context of an initial study. Nevertheless, in order to derive a second general requirement for the performance of FC, now bound to the transfer time, it is possible to analyze the derivative of the function \ref{e18} only at the point $\varepsilon=1$. Thus, at this point, the derivative of the function \ref{e18} is given by,

\begin{equation}
\label{e19}
\frac{d\tau}{d\varepsilon}\bigg|_{\varepsilon = 1} = \frac{1 - e^\nu}{\lambda} + \nu e^\nu \left( \frac{2 - \lambda}{2 \lambda} \right).
\end{equation}

If $d\Delta t/d \varepsilon > 0$ at this point, it is possible to imply that function will continue to grow, and the proposed system is most likely non-applicable. Thereby, it is necessary to ensure that at least on the initial point the function has a declining profile. Analyzing the equation \ref{e19}, it is possible then to guarantee a negative inclination when,

\begin{equation}
\label{e20}
\lambda > 2 \left( 1 - \frac{1 - e^{-\nu}}{\nu} \right).
\end{equation}

Comparing the obtained condition with the expression \ref{e14}, it is possible to observe that the values are similar, but corrected by a factor influenced mainly by the $\Delta V$ of the maneuver and performance characteristics of the used thruster. It is important to note that the obtained requirement does not provide a guarantee that the application of FCs would be advantageous for the mission, but rather provides a general lower boundary for the performance of the system. On the contrary of the previous requirement, in expression \ref{e14}, the present one depends on the mission profile, by the $\Delta V$. 

In order to increase the fidelity of the estimation it is possible to include the growth of the power system and FC mass with the increase of total power. Considering a linear variation of the mass of both system and that $\alpha_{SC}$ and $\alpha_{FC}$ are the specific power of each of them respectively, it is possible to final to define the dry mass of the spacecraft as,

\begin{equation}
m_f = m_u + \alpha_{SC} P_{SC} \varepsilon + \alpha_{FC} P_{SC} (\varepsilon - 1),
\end{equation}

where $m_u$ is the useful mass.  Using this definition with equation \ref{e11}, it is possible to define a corrected payload mass ratio equation that expresses the amount of dry mass that can be used for other purposes than power and fuel cell systems,

\begin{equation}
\frac{m_u}{m_0} = \exp \left( - \frac{\Delta V}{u_q (\varepsilon)} \right) - \frac{ \alpha_{SC} P_{SC} \varepsilon + \alpha_{FC} P_{SC} (\varepsilon - 1)}{m_0}.
\end{equation}

Using a similar approach of the derivation of the condition \ref{e20}, it is possible to define a requirement for the specific power values that ensures an increasing profile of the payload ratio for a given $1 \leq \varepsilon < \varepsilon_m$,

\begin{equation}
\frac{d}{d\varepsilon} \left( \frac{m_u}{m_0} \right)\bigg|_{\varepsilon = \varepsilon_m} = \left( \frac{\nu}{2 \varepsilon_m^{3/2}} \left(1 - \frac{\varepsilon_m + 1}{\lambda}\right) - \phi \right)\frac{m_f}{m_0} = 0,
\end{equation}

where $\phi=(\alpha_{SC}+\alpha_{FC} ) P_{SC}/m_f$. Giving the following condition to ensure that the extremum of the function is at $\varepsilon_m$,

\begin{equation}
\lambda = \frac{\nu (\varepsilon_m + 1)}{\nu - 2\varepsilon_m^{3/2}\phi}.
\end{equation}

From this relation it is straightforward to observe that, to ensure that it is possible to select specific powers with positive values, one must have a positive value in the right hand side. One way to ensure this is having $k_c > (\varepsilon_m+1) P_{SC}/\dot{m}_p$, which shows close concordance with the condition expressed in \ref{e14}. Since the definition of useful mass ratio does not modify the function of payload ratio in equation \ref{e11}, the expression of the maneuver time continues to hold true.

\subsection{Case 2: product of fuel cell as a propellant}

In this case, it is considered that the propellant fed to the thruster is composed by the exhaust of the fuel cell, $\dot{m}_{FC}$, and a separate feeding system, $\dot{m}_{p}$. In the first part we study a combination of both flows in order to determine the impact of adding a fuel cell to a given propulsion system. Next, in the second part, we consider the propellant to be exclusively the exhaust of the fuel cell. 

It is important to note that output product substance may not be directly compatible to an electric thruster, nevertheless several works in literature demonstrate the usage of common reaction products, such as water and carbon dioxide, as a propellant for different types of thruster. We leave for a future work the consideration of the performance loss due to the usage of alternative propellants.

Taking into consideration again the definition of thrust efficiency, the exhaust of the system can be simply calculated as,

\begin{equation}
u_e = \sqrt{\frac{2 \eta_TP_{SC}}{\dot{m}_T} + \frac{2 \eta_T k_c \dot{m}_{FC}}{\dot{m}_T}}\,,
\label{eq:ue_usep}
\end{equation}

where $\dot{m}_{T}$ is the total propellant mass flow rate, given by $\dot{m}_{T} = \dot{m}_p + \dot{m}_{FC}$. It is interesting to not that the first term in the square root represents a ``solar electric propulsion'' part and the second term is relative to a ``chemical propulsion'' part. For $P_{SC} = 0$ or very high $\dot{m}_{FC}$, the expression becomes very similar to an ideal chemical thruster, and, if it is considered that $\dot{m}_{T} = \dot{m}_{FC}$ we obtain

\begin{equation}
	u_e = \sqrt{2 \eta_T k_c},
\end{equation}

which holds some resemblance with the expression for the thermal acceleration $u_e \approx \sqrt{2c_pT}$ demonstrated by Jahn\cite{jah64}. Comparing this expression with equation \ref{eq:uqm} it is clear that, as expected, using the fuel cell exhaust as propellant instead of discarding it, a specific impulse 2 times higher is obtained. 

 When $k_c=0$ the relations are reduced to the equations of a common electric thruster. The specific impulse and the thrust of a propulsion system in this configuration will always be higher than a system using only solar power, since the energy from the chemical reaction is added to the thrust.

Considering a similar approach to case 1, it is possible to define again $P_T = \varepsilon P_{SC}$ and $\dot{m}_{FC} = (\varepsilon - 1)P_{SC}/k_c$, and rewrite equation \ref{eq:ue_usep} in the form,

\begin{equation}
	u_e = \sqrt{\frac{2 \eta_T \varepsilon P_{SC}}{\dot{m}_p + (\varepsilon - 1)\frac{P_{SC}}{k_c}}}.
\end{equation}

Defining once more the normalized variables $\lambda = k_c\dot{m}_p/P_{SC}$ and $\chi = u_e/u_{SEP}$ it is possible to write the exhaust velocity expression as,

\begin{equation}
\label{eq:chi_2}
	\chi = \sqrt{\frac{\lambda \varepsilon}{\lambda + \varepsilon - 1}}.
\end{equation}

Noting the resemblance between equations \ref{eq:chi_2} and \ref{e10} it is possible to observe that both architectures have a similar behavior. However, in contrast to the first case, equation \ref{eq:chi_2} does not present any extremum points making $\chi$ increase or decrease monotonically with $\varepsilon$. To guarantee that $\chi$ always increases we can analyze its derivative in the form, 

\begin{equation}
	\frac{d\chi}{d\varepsilon} = \frac{(\lambda - 1)\lambda}{2 \chi (\lambda + \varepsilon - 1)^2}.
\end{equation}

Which then imposes a restriction for the fuel cell given by $\lambda > 1$, or,

\begin{equation}
	k_c > \frac{P_{SC}}{\dot{m}_p} = \frac{u_{SEP}^2}{2\eta_T}.
	\label{eq:restriction2}
\end{equation}

It is possible to note again the similarity between the restrictions \ref{eq:restriction2} and \ref{e14}, but in the present scenario $k_c$ has a more relaxed constraint, which is two times higher than in the first case. Of course, this can be justified by the fact that the fuel cell flow is actually being used to give kinetic energy for the spacecraft, and \ref{eq:restriction2} is only requiring that the energy in the fuel cell flow is higher than the energy contained in the ``solar electric propulsion'' flow, given by $P_{SC}/\dot{m}_p$.

We assume now that the thruster uses exclusively the products of the fuel cell as propellant, with no external feeding system. To this end, it is assumed that the total propellant flow and the fuel cell exhaust is given by $\dot{m}_p$, i.e. $\dot{m}_T = \dot{m}_p$, $\dot{m}_{FC} = \dot{m}_p$. Substituting these definitions and dividing equation \ref{eq:ue_usep} by $u_{SEP}$ it is straightforward to define

\begin{equation}
\chi = \sqrt{\frac{P_{SC} + k_c \dot{m}_{p}}{P_{SC}}} = \sqrt{1 + \lambda}.
\end{equation}

Noting that a normalized thrust is simply given by $\xi=\chi$, a relation between specific impulse and thrust can be written as

\begin{equation}
\chi = \frac{1 + \lambda}{\xi},
\end{equation}

resembling the common relation of power, thrust and specific impulse. It is interesting to observe here that, as expected, if the propellant is only composed by the reaction products, the trend of the specific impulse is always crescent, once we are adding energy to the flow without any drawback.

\begin{figure}

  \centering
    \includegraphics[width=0.6\textwidth]{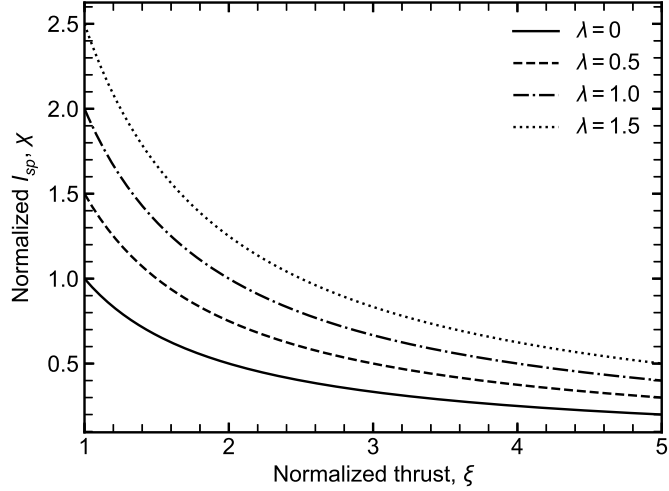}
  
  \caption{Normalized specific impulse in function of the normalized thrust for different values of $\lambda$}
\label{fig3}
\end{figure}

Figure \ref{fig3} plots the equation for several values of $\lambda$. It is possible to observe that the energy addition from the chemical reaction increases the overall power when $\lambda$ increases.  

Taking into consideration the mass of power supplies that increases with power, as in the first case, it is possible to derive a corrected payload ratio,

\begin{equation}
\frac{m_u}{m_0} = \exp \left(- \frac{\nu}{\sqrt{1 + \lambda}} \right) - \frac{\alpha_{FC} P_{SC}}{m_0} \lambda - \frac{\alpha_{SC} P_{SC}}{m_0}
\label{eq:payload},
\end{equation}

where $\alpha_{FC}$ and $\alpha_{SC}$ are the specific masses of the fuel cell and the other power supplies in kg/W, respectively. Simply differentiating the obtained equation about $\lambda$ it is possible to find an extremum point which characterizes the optimum specific energy for a given mission. This expression is found as
 
\begin{equation}
\frac{d}{d\lambda} \left( \frac{m_u}{m_0} \right) = \left( \frac{\nu}{2 (1 + \lambda_m)^{3/2}} - \phi_{FC} \right) \frac{m_f}{m_0} = 0 \,;
\end{equation}

\begin{equation}
\lambda_m = \left( \frac{\nu}{2\phi_{FC}} \right)^{2/3} - 1,
\end{equation}

where $\phi_{FC}=(\alpha_{FC} P_{SC})/m_f$. Giving the condition,

\begin{equation}
k_c < \frac{1}{2} \left( \frac{m_f}{\dot{m}_p} \frac{\Delta V}{\alpha_{FC} \sqrt{\eta_T}} \right)^{2/3} - \frac{P_{SC}}{\dot{m}_p}.
\end{equation}

Meaning that if $k_c$ is superior to this specified value the payload ratio will start to decrease since the fuel cell mass will get proportionally higher. Thus if the performance of the fuel cell is increased, with the decrement of $\alpha_{FC}$, a wider range of values for $k_c$ are permitted to increase the performance of the mission. For the applicability of this scheme it is important then to guarantee that $\lambda_m$ is positive, 

\begin{gather}
	\lambda_m = \left( \frac{\nu}{2\phi_{FC}} \right)^{2/3} - 1 > 0 \,;\\
	\phi_{FC}  < \frac{\nu}{2}.
\end{gather}

Yielding the condition for the specific mass of the fuel cell in the form

\begin{equation}
	\alpha_{FC} < \frac{m_f \Delta V }{2 u_{SEP} P_{SC}}. 
\end{equation}

It is important to note that even after the calculated limit, if $\lambda_m$ is positive, there is a region where the resultant payload ratio is still higher than the payload ratio without the fuel cell action. It is possible to calculate an approximate limit where it is still advantageous the usage of the fuel cell. We can then divide the equation \ref{eq:payload} by itself with $\lambda = 0$ and obtain

\begin{equation}
	\left[ 1 - \frac{\alpha_{FC} P_{SC}}{m_f} \right] \exp\left(\nu \left( 1 - \frac{1}{\sqrt{1 + \lambda}} \right)\right) = 1.
\end{equation}

Where it is imposed that this ratio should be equal the unity to represent the limits to the region where the performance of the system is higher than the case without fuel cell. Using a Taylor expansion to represent the exponential part, and truncating it in the second order, we can achieve the approximate value

\begin{equation}
	\lambda_{max} \approx \frac{4\nu - \phi_{FC}}{\nu(\nu + 3)}.
\end{equation}

Then, if $\lambda_m > 0$, the region between $\lambda = 0$ and $\lambda = \lambda_{max}$ always increases the payload ratio of the mission.

\section{Conclusion}
This work presented critical study on the feasibility of using fuel cell systems to modify the performance of electric thrusters using just basic relations of propulsion performance. In the introduction section we presented a background discussion showing the main points about history of fuel cells and their usage on spaceflight. In the first section we showed a basic modeling procedure for the fuel cell performance based on the chemical energy contained in the used reactants. 

Using this basic fuel cell model, in the second section, we begin a study of cases, the first consisting of a spacecraft using separate propellant feeding systems for the propulsion and fuel cell systems, and the second consisting of a thruster using the reaction product of the fuel cell as its propellant. In the first scenario it is showed that if the fuel cell product is simply discarded, as expected, the overall performance of the propulsion system is bound to decrease unless we possess reactants that yield energy levels superior to a quantity denoted by $k_0 = 2P_{SC}/\dot{m}_p=u_{SEP}^2/\eta_T$. Considering then the performance of the hydrogen-oxygen fuel cell showed in table \ref{t1} as the maximum attainable, in order to achieve an increment in performance in this case, the thruster should have less than $360 \, \mathrm s$ of specific impulse, proving the non-applicability of this scheme for the vast majority of the electrical thrusters available today. It is also showed in this section that if no power is available from external sources, like solar panels, the performance of an electrical thruster will be always poorer than an chemical thruster with the same propellants, simply because of the relation between enthalpy and free energy of formation.

In the next case, where it was considered that the reaction products were used as propellant, it is shown that both the specific impulse and the thrust increase for any given $k_c$, since we are simply adding energy to the system without any apparent drawback. Next we considered the impact of the fuel cell mass in the spacecraft performance and this imposed a restriction for the applicability of this scheme considering the mission $\Delta V$ and the specific mass of the fuel cell. In order to have an increment in performance it is shown that the fuel cell must at least meet the requirement calculated as $\alpha_{FC} < m_f \Delta V  / 2 u_{SEP} P_{SC}$. Lastly it was shown if lambda is inside a certain range between $0$ and $\lambda_{max}$ while $\lambda_m$ is positive, the payload ratio of the mission increases.

\section*{Acknowlegments}

ISF thanks CNPq,
project number PDE(234529/2014-08), and also FAPDF project number 0193.000868/2015,  call 03/2015.

\bibliographystyle{ieeetr}
\bibliography{bibliography.bib}

\newpage
\appendix
\section*{List of symbols}

\begin{tabular}{ll}
$E^0$           & Maximum expected voltage, V                                                            \\
$E_c$           & Voltage in a single cell, V                                                            \\
$F$             & Faraday constant, $ \approx 96485.33$ sA/mol                                           \\
$I$             & Current produced by a fuel cell, A                                                     \\
$I_{sp}$        & Specific impulse, s				                                                     \\
$k_{c}$        	& Specific energy, J/kg       			                                                 \\
$k_{0}$        	& Specific energy limit given by $2P_{SC}/\dot{m}_p$, J/kg       	                     \\
$M$             & Molar mass, kg/mol                                                                     \\
$M_{FC}$        & Molar mass of the fuel cell product, kg/mol                                            \\
$m_0$           & Initial mass of the spacecraft, kg                                                     \\
$m_f$           & Dry mass of the spacecraft, kg                                                         \\
$\dot{m}_{FC}$  & Output mass flow rate of a fuel cell, kg/s                                             \\
$m_q$           & Total mass of equivalent propellant $\dot{m}_p+\dot{m}_{FC}$, kg                       \\
$\dot{m}_p$     & Propellant mass flow rate, kg/s                                                        \\
$m_u$           & Useful mass (dry mass excluding power systems masses), kg                              \\
$n$             & Number of equivalent electrons produced per mole                                       \\
$P$             & Power, W                                                                               \\
$P_{FC}$        & Power generated by the fuel cell, W                                                    \\
$P_{SC}$        & Power generated by the solar cells, W                                           \\
$P_T$           & Total power, W                                                                         \\
$T$             & Thrust, N                                                                              \\
$T_q$           & Equivalent thrust, N                                                                   \\
$u_e$           & Real exhaust velocity, m/s                                                             \\
$u_{ch,m}$      & Theoretical maximum of the exhaust velocity in a chemical thruster, m/s                \\
$u_q$           & Equivalent exhaust velocity, m/s                                                       \\
$u_{SEP}$       & Real exhaust velocity when $P_T=P_{SC}$, W                                             \\
$\alpha_{SC}$   & Specific power of the solar power system, kg/W                                         \\
$\alpha_{FC}$   & Specific power of the fuel cell power system, kg/W                                     \\
$\beta$         & Non-dimensional mass flow, $\beta = \dot{m}_p/\dot{m}_{SEP}$                           \\
$\Delta G^0$    & Standard Gibbs free energy of formation, J/mol                                         \\
$\Delta H^0$    & Standard enthalpy of formation, J/mol                                                  \\
$\Delta V$      & Delta-V of the mission, m/s                                                            \\
$\Delta t$      & Mission time, s                                                                        \\
$\varepsilon$   & Ratio of total power to solar cell power, $\varepsilon=P_T/P_{SC}$                     \\
$\varepsilon_m$ & Power ratio that guarantees maximum payload ratio                                      \\
$\eta_T$        & Thrust efficiency                                                                      \\
$\eta_V$        & Voltage efficiency, $\eta_V=E_c/E^0$                                                   \\
$\lambda$       & Ratio of power deposited with FC and SC, $\lambda=k_c \dot{m}_p/P_{SC}$                \\
$\nu$           & Non-dimensional V, $\nu=\Delta V/u_{SEP}$                                              \\
$\xi$           & Non-dimensional thrust, $\xi=T/\dot{m}_{SEP} u_{SEP}$                                  \\
$\phi$          & Non-dimensional power source mass, $\phi=(\alpha_{SC}+\alpha_{FC} ) P_{SC}/m_f$        \\
$\phi_{FC}$     & Non-dimensional power source mass of the fuel cell, $\phi_{FC}=\alpha_{FC} P_{SC}/m_f$ \\
$\tau$          & Non-dimensional time, $\tau=\Delta t \dot{m}_p/m_f$                                    \\
$\chi$          & Non-dimensional exhaust velocity, $\chi=u_q/\sqrt{2 \eta_T P_{SC}/\dot{m}_p}$          \\
$\chi'$         & Non-dimensional exhaust velocity, $\chi=u_e/u_{SEP}$                                  
\end{tabular}

\end{document}